\documentclass[a4paper,12pt]{article}
\usepackage{epsfig}
\usepackage{citesort}
\date{\today}
 \normalsize

\newcommand{\bmat}{\left(\begin{array}}
\newcommand{\emat}{\end{array}\right)}
\newcommand{\be}{\begin{equation}}
\newcommand{\ee}{\end{equation}}
\newcommand{\bea}{\begin{eqnarray}}
\newcommand{\eea}{\end{eqnarray}}
\def\ie{{\it i.e.}}
\def\susy{\mbox{\tiny SUSY}}
\def\sm{\mbox{\tiny SM}}
\def\lsim{\raise0.3ex\hbox{$\;<$\kern-0.75em\raise-1.1ex\hbox{$\sim\;$}}}
\def\gsim{\raise0.3ex\hbox{$\;>$\kern-0.75em\raise-1.1ex\hbox{$\sim\;$}}}
\def\Frac#1#2{\frac{\displaystyle{#1}}{\displaystyle{#2}}}

\begin{document}
\renewcommand{\thefootnote}{\fnsymbol{footnote}}
\rightline{IPPP/03/36} \rightline{DCPT/03/72}
\vspace{.3cm} 
{\large
\begin{center}
{\bf Can the CKM phase be the only source of CP violation}
\end{center}}
\vspace{.3cm}

\begin{center}
Shaaban Khalil$^{1,2}$ and Veronica Sanz$^{1}$\\
\vspace{.3cm}
$^1$\emph{IPPP, Physics Department, Durham University, DH1 3LE,
Durham,~~U.~K.}
\\
$^2$ \emph{Ain Shams University, Faculty of Science, Cairo, 11566,
Egypt.}

\end{center}

\vspace{.3cm}
\hrule \vskip 0.3cm
\begin{center}
\small{\bf Abstract}\\[3mm]
\end{center}
We address the question of whether the CP violating phase in the CKM mixing matrix 
($\delta_{\mathrm{CKM}}$) can be the only source of all CP violation. We show that 
in supersymmetric models with new flavour structure beyond the Yukawa matrices, 
$\delta_{\mathrm{CKM}}$ can generate the required baryon asymmetry
and also accounts for all observed results in $K$ and $B$ systems, 
in particular $\varepsilon'/\varepsilon$ and the CP asymmetry of 
$B \to \phi K_S$. 

\begin{minipage}[h]{14.0cm}
\end{minipage}
\vskip 0.3cm \hrule \vskip 0.5cm
%
\section{{\bf Introduction}}

In the Standard Model (SM), the only source of CP violation arises from the Cabibbo-
Kobayashi-Maskawa (CKM) phase which, so far, can saturate the observed  
CP violation in the kaon system. However, it has been pointed out that 
the strength of the CP violation in the  SM is not enough to generalize the cosmological 
baryon asymmetry of our universe and a new source of CP violation is required 
\cite{ewSM}. Also, the recent experimental 
results on the CP asymmetries of $B \to \phi K_S$ ($S_{\phi K_S}$) \cite{BKphi} and 
$B \to \eta' K_S$ ($S_{\eta' K_S}$) \cite{BKeta} reveal deviations from the SM 
prediction which have been considered as  a hint of a new CP violating source beyond 
the SM. In this paper we demonstrate that this conclusion is not necessarily true. 
We argue that in the presence of new flavor structures beyond the Yukawa matrices, 
the CP violating phase in the CKM mixing matrix can accommodate 
all the observed results, in particular $\varepsilon'/\varepsilon$, $S_{\phi K_S}$ 
and the cosmological baryon asymmetry, which might be different from
the SM predictions. In fact, supersymmetric models with additional 
sources of CP violation usually face severe constraints from the experimental bounds 
on the electric dipole moments (EDM) of the neutron, the electron and the mercury atom 
\cite{Abel:2001vy}, leading to what is known as SUSY CP problem.

It is customary assumed that the supersymmetric extensions of the SM
have additional sources of CP violation which may arise from the complexity
of the soft SUSY breaking and also from the SUSY conserving
$\mu$-parameter. However, the SUSY breaking and the CP
violation are in general not related and they could have different
origins and different scales of breaking. Indeed, in a broad class of
string and brane models, it is feasible to spontaneously violate CP
 and generate the CKM phase while the SUSY parameters
remain real \cite{Khalil:2001dr}. Therefore, we assume that the SUSY
breaking sector conserves CP and the only source of CP violation is
$\delta_{\mathrm{CKM}}$. The details of the realization of this
scenario in supergravity inspired models derived from heterotic and
type I string theories will be presented elsewhere.

It is important to note that even in case of real SUSY soft breaking
terms, the EDM experimental bounds impose strong constraints on the
flavour structure of the trilinear parameters \cite{Abel:2001cv}.
These constraints are very sensitive to the structure of the Yukawa
couplings. It has been shown that the constraints on the $A$-terms are
stronger in the case of non hierarchical Yukawas than in the hierarchical case.
  However, there are a number of possible flavour
patterns that can overcome these constraints and suppress SUSY
contributions to the EDMs by many orders of magnitude.

This paper is organized as follows. In section 2 we present the
non-minimal flavour SUSY models that we will use.  In section 3 we
show that within this class of models one can generate enough baryon
asymmetry in the Universe with only the $\delta_{\mathrm{CKM}}$
phase. Section 4 is devoted to the SUSY prediction of the direct CP
violation parameter $\varepsilon'/\varepsilon$. In section 5 we
analyze the CP violation in the $B$-sector and show that with the new
source of flavour, it is natural to reduce $S_{\phi K_S}$ to
negative values inside the experimental range. Our conclusions are
presented in section 5.


\section{{\bf{SUSY models with non-minimal flavour violation}}}
As mentioned in the introduction, non-universal soft breaking terms
are the crucial ingredients in order to have a new flavour structure beyond
the usual Yukawa matrices and then enhance the effect of the
$\delta_{CKM}$ phase. Moreover, most of string inspired models
naturally lead to SUSY realizations with non-universality, in particular
non-universal $A$-terms and non-universal scalar masses for the  quark
singlets $M^2_{\tilde{u}}$ and $M^2_{\tilde{d}}$.  Concerning the
non-universality of the $A$-terms, it has been emphasized
\cite{previous} that in this case the gluino contributions to the CP
violating processes are enhanced through large imaginary parts of the
$LR$ mass insertions. However, the  observed EDM
bounds restrict the non-universality of even real $A$-terms. In
particular, the EDM of the mercury atom implies that
$\mathrm{Im}(\delta_{11}^{d(u)})_{LR} \lsim 10^{-7}-10^{-8}$ and
$\mathrm{Im}(\delta_{22}^{d})_{LR} \lsim 10^{-5}-10^{-6}$
\cite{Abel:2001vy}, which are strong constraints for most of the SUSY
models. Thus, only certain type of patterns for $A$-terms can be allowed.

We assume that the trilinear couplings have the following structure:
\begin{equation}
A^{(d,u)} = m_0 \left( \begin{array}{ccc}
a^{(d,u)} & b^{(d,u)} & c^{(d,u)} \\
a^{(d,u)} & b^{(d,u)} & c^{(d,u)}  \\
a^{(d,u)} & b^{(d,u)} & c^{(d,u)}
\end{array} \right) \;.
\label{A-terms}
\end{equation}
This pattern is known as factorizable $A$-terms. Since $A_{ij} Y_{ij}= 
Y. \mathrm{diag}\left(a,b,c\right)$,
the relevant mass insertions for the EDM contributions $(\delta^{(d,u)}_{11})_{LR}$ are
suppressed by the factors $m^{(d,u)}/\tilde{m}$. This special structure of $A$-terms 
could arise naturally in D-brane models as explained in Ref.\cite{Khalil:2000ci}. We have 
explicitly checked that in this case the EDMs are many orders of magnitude 
below the experimental 
limits for all possible values of the parameters $a,b$ and $c$. 

Another possibility for new flavour structure in supersymmetric models is to have 
non-universal squark masses. While the masses of the squark doublets are 
bounded by $\Delta M_K$ and $\varepsilon_K$ to be almost universal, the masses 
of the squark singlets are essentially unconstrained. The non-universality of the 
squark singlets can be also obtained in string and D-brane models.
However, as we will show, we do not need to consider this type of non-universality 
and it will suffice to use $A$-terms as in Eq.(\ref{A-terms}) in order to satisfy all 
the CP violating measurements. Hence thorough the paper 
we will assume that at GUT scale the soft scalar
masses and the gaugino masses are universal and are given by $m_0$ and $m_{1/2}$
respectively.

It is worth mentioning that with non-universal soft SUSY breaking terms, the Yukawa 
textures play a crucial rules in the CP and flavour supersymmetric results and one 
has to specify the type of the  employed Yukawa in order to  completely define the model. 
In our analysis, we assume that the Yukawa has the non-hierarchical structure used 
in Ref.\cite{Abel:2001cv}.
%

\section{{\bf{Baryon asymmetry and non-universal soft terms}}}

The most precise measurement of the ratio of the baryon-to-entropy
in the Universe is given by \cite{WMAP}
\be
0.78\times 10^{-10} <\eta=\frac{n_B}{s} < 1.0\times 10^{-10} \ ,
\ee
where we require that $\eta$ lies on the 3$\sigma$ range. 

As mentioned in the introduction, in the context of the SM there is no
mechanism able to produce enough amount of BAU at the electroweak
phase transition \cite{ewSM} and several models with additional
sources of CP violation beyond the SM phase $\delta_{CKM}$ have been
proposed, in particular supersymmetric models are well motivated
extensions of the SM mechanism \cite{oldMSSM,ewMSSM,Cohen:1994ss}.
However the experimental bounds on the EDM of the electron, neutron
and mercury impose severe constraints on these additional sources of CP
violation. In this section we show that in supersymmetric theories
with non-universal soft terms the electroweak baryogenesis can take
place with no CP phases besides the $\delta_{\mathrm{CKM}}$.

The first step in electroweak baryogenesis is to identify which local
charges are approximately conserved in the symmetric phase
\cite{Cohen:1994ss}. Once a charge asymmetry is produced on the wall
separating both phases, they will efficiently diffuse to the unbroken
phase, where sphaleron processes are active, and a net baryon number
will be produced \cite{ewMSSM}. In the supersymmetric theories, these
charges are the axial stop charge and the Higgsino charge. The latter
had received much of the attention mainly because the left handed
squarks were assumed to be very heavy and in this case stop current is
suppressed compared with the Higgsino current. However, the Higgsino
contributions depend on the relative phase ($\phi_{\mu}$) between the
$\mu$ parameter and the gaugino mass $M_2$ which is strongly
constrained by the EDM limits. It has been emphasized in
Ref.\cite{Delepine:2002as} that in SUSY models with new flavour
structure beyond the usual Yukawa matrices, the stop contributions are
enhanced and can accommodate the measurement of the  BAU. In our model with no
new SUSY CP violating phases, the stop contribution is found to be
significant and can easily provide enough baryon asymmetry.

Following the notation of Refs.\cite{ewMSSM,Delepine:2002as}, the
right-handed squark contributions to the baryon asymmetry is given by
\bea \frac{n_B}{s} \approx 10^{-9} \times \, \mathrm{Im}\left( \mu~
\mathrm{Tr}\{ [(Z_R (Y^A_u)^T W_L^{\dagger}) \,I_{RR} ]\, \ .  [ W_L
h_u^* Z_R^{\dagger}] \} \right).  \eea
The prefactor is obtained from: \bea -\frac{v(T)^2 \Delta \beta}{T^3}
\, \frac{5 \Gamma_{ws} }{96 \pi^2 g_{*}} \, \frac{v_w}{D R+v_w^2} \,
\frac{ f_{\lambda_{+}}}{\lambda_{+}} \, \frac{ (9 k_T-k_B) (9 k_Q
k_B-8 k_B k_T-5 k_B k_Q)}{k_T-k_B} \eea where all the definitions can
be found in Ref.\cite{Delepine:2002as}.  The matrix $I_{RR}$ is the integration
corresponding to the propagators of the right and left-handed squarks
and $W_L$ ($Z_R$) is the diagonalizing matrix of the $LL$ ($RR$)
squark squared mass matrix. The matrix $Y^A_u$ is given by
$(Y^A_u)_{ij}=(Y_u)_{ij} A^u_{ij}$.  
In order to estimate a model independent bound on the relevant $LR$
mass insertions, it was assumed in  Ref. \cite{Delepine:2002as} that the 
squark mass matrices
$M_{LL}^2$ and $M_{RR}$ were diagonal, \ie, $W_L=Z_R = 1$. Hence, the
typical form of the baryon-to-entropy ratio was given by \bea
\frac{n_B}{s} \approx 10^{-9} \, I_{RR} \, \,\mu Y_t^2 \frac{\langle
m_{\tilde{q}}^2\rangle}{m_t} \mathrm{Im}(\delta_{LR}^u)^*_{3i} \eea
which imposes strong bounds on the
$\mathrm{Im}(\delta_{LR}^u)^*_{3i}$, typically to be of the order of
${\cal O}(10^{-1})$. Such values might not be easy to obtain in SUSY models and
also it may contradict the constraints derived from the experimental
measurements of $B-\bar{B}$ mixing and the CP asymmetry in the decay of $B \to
J/\psi K_S$ \cite{Gabrielli:2002fr}. In our analysis, we find that due to the
non-universality of the soft terms, the diagonalizing matrices $W_L$
and $Z_R$ turn out to have sizable phases (${\cal O}(1-10^{-1})$), playing an important
role in relaxing the constraints on $\mathrm{Im}(\delta_{LR}^u)^*_{3i}$. 
\begin{figure}[ht]
\begin{center}
\hspace*{-7mm}
\epsfig{file=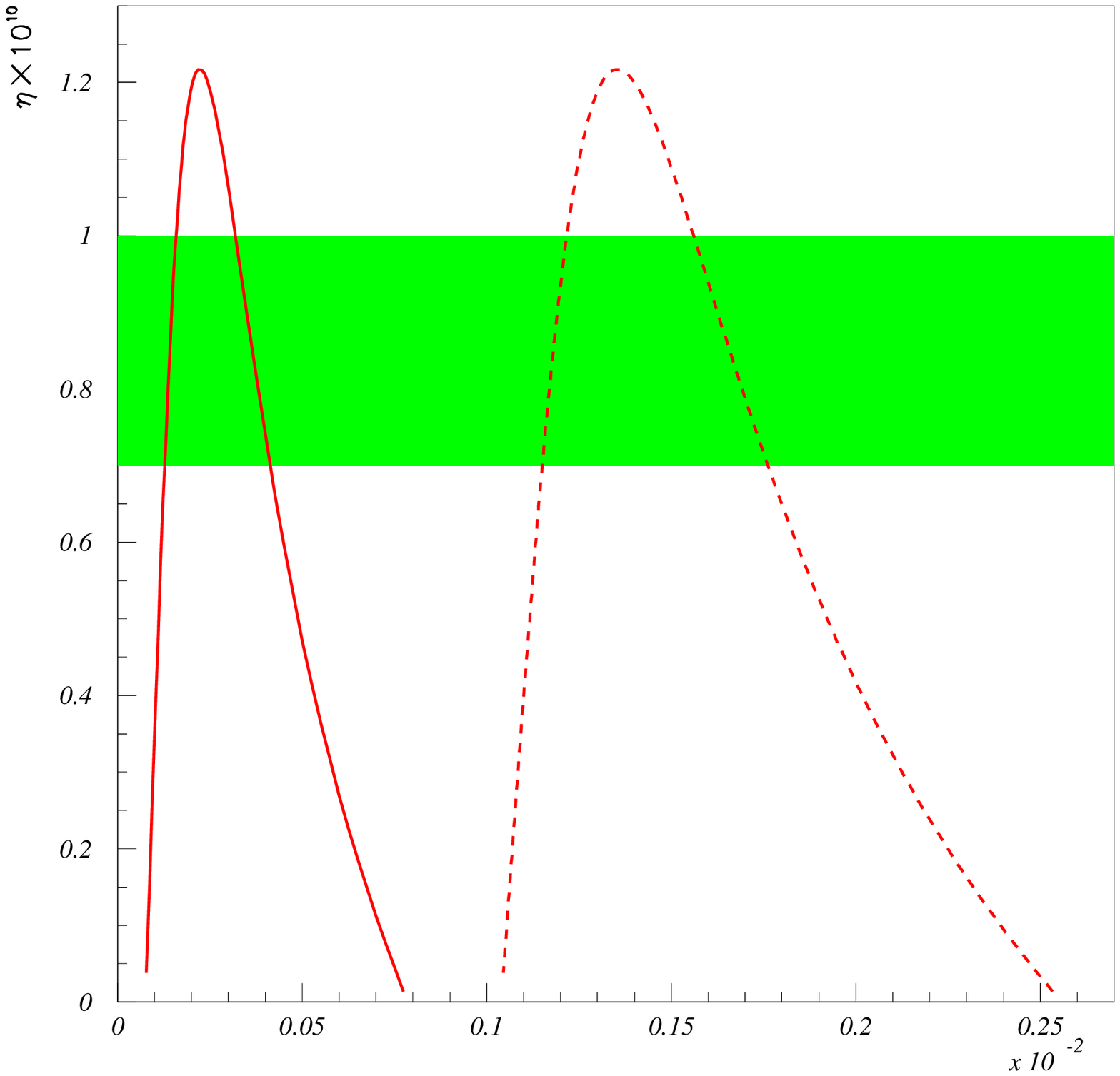,width=.5\textwidth,height=.25\textheight}\quad
\epsfig{file=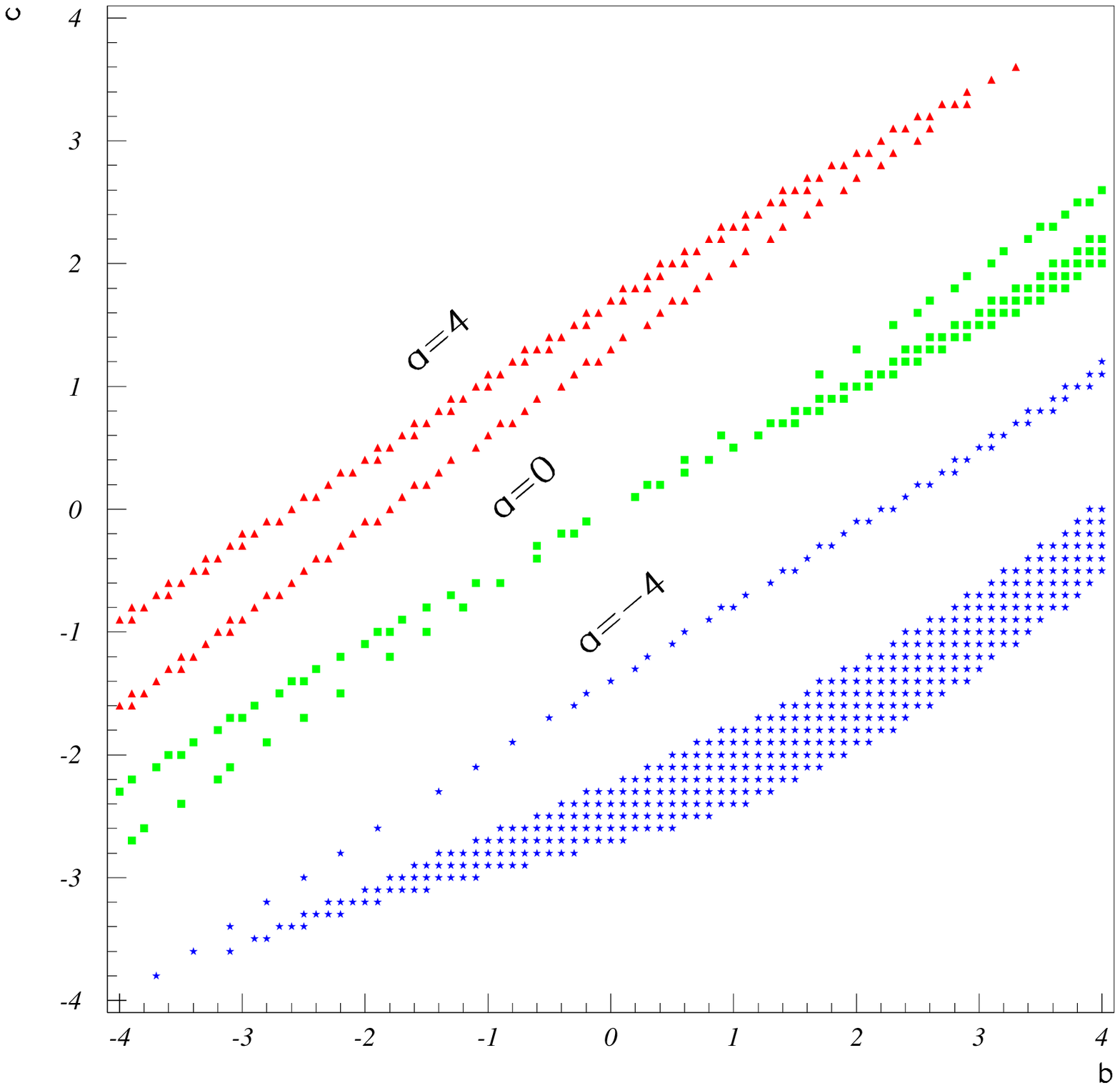,width=.5\textwidth,height=.25\textheight}\\
\caption{\footnotesize{(Left) The BAU as functions of the 
$\mathrm{Im}(\delta^u_{LR})_{31}$ 
and  $\mathrm{Im}(\delta^u_{LR})_{32}$ for $m_0=m_{1/2}=200$ GeV. (Right) The allowed
range of the trilinear couplings by BAU limits.}}
\label{fig1}
\end{center}
\end{figure}

In   Fig. 1, we plot the amount of baryon asymmetry in terms of the
imaginary part of the relevant
mass insertions: $(\delta^u_{LR})_{31}$ and $(\delta^u_{LR})_{32}$ for
$m_0=m_{1/2}=200$ GeV,
$a^u=4$, $c^u=3$ and $b^u$ varying in the range $b^u=(1.5-2.7)$.
As can been seen from this figure,  the typical values of the mass
insertions lie on a scale 2 or 3 orders
of magnitude smaller than the estimates given in
\cite{Delepine:2002as}.
Also
we present in this figure the regions for the trilinear couplings
$a^u$, $b^u$ and $c^u$ that lead to $n_B/s$ within the observed range.

\section{{\bf{Supersymmetric contributions to meson decays }}}

The enhancement of CP violation by the new flavor structure has also
interesting consequences in kaon and B-systems.  
In particular, we will study the effect on  
the direct CP violation in the kaon decays $\varepsilon'/\varepsilon$ and
the mixing CP asymmetry of the $B\to \phi K_S$ process. The measured
values of the indirect CP violation $\varepsilon_K$ and the CP asymmetry in $B\to
J/\psi K_s$ are saturated by the SM contributions with $\delta_{CKM}$ of
order one. It is also worth mentioning that due to the fact that BAU 
constrains only the up-squark sector and has no impact on the down-squark
sector, the dominant gluino contributions to the $\varepsilon'/\varepsilon$ and $S_{\phi
K_S}$ are free from the constraints mentioned in the previous section,
however for the chargino contribution it should be taken into account.
We find that the chargino loops are irrelevant to the computed CP violating observables.

\begin{figure}[ht!]
\begin{center}
\hspace*{-7mm}
\epsfig{file=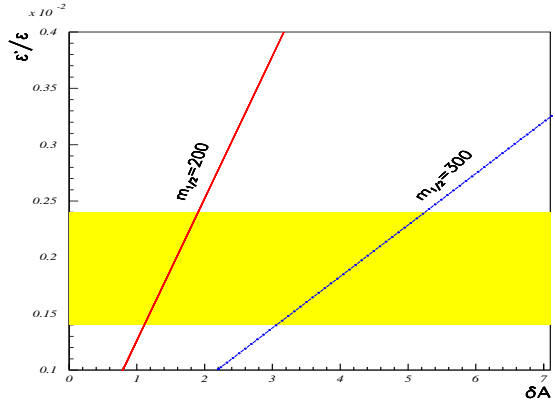,width=.5\textwidth,height=.25\textheight}\\
\caption{\footnotesize{$\varepsilon'/\varepsilon$ versus $\delta A$
 for $\tan\beta=5$ and  $m_0=200$ GeV . 
 The red (solid) line corresponds to  $m_{1/2}=200$ GeV, while the
blue (dashed) line corresponds to 
$m_{1/2}=300$ GeV. 
The yellow band corresponds to the experimental 1 $\sigma$ uncertainty. }  }
\label{fig2}
\end{center}
\end{figure}

Direct CP violation has been observed  in kaon decays and its measured value lies in the range 
\be
Re \left(\frac{\varepsilon'}{\varepsilon}\right)=(1.8 \pm 0.4) \times 10^{-3}
\ee
the SM contribution \cite{Buras:1993dy} is dominated by the QCD
penguin $Q_6$ and the electroweak penguin
$Q_8$ density operators. It can be written as 
\be
Re \left(\frac{\varepsilon'}{\varepsilon} \right)^{SM}=\frac{Im(\lambda_t
\lambda_u^*)}{\lambda_u}  F_{\varepsilon'}
\ee
where $F_{\varepsilon'}$ depends on the non-perturbative parameters
$B_6^{(1/2)}$ and $B_8^{(3/2)}$ associated with the QCD and
electroweak operators, respectively
\be
 F_{\varepsilon'}=\left( \alpha_1 B_6^{(1/2)} - \alpha_2 B_8^{(3/2)}\right) \ .
\ee
The values of $\alpha_1$, $\alpha_2$ are given in \cite{Buras}.  The computation
of $F_{\varepsilon}$ requires an accurate determination of the involved
penguin operators and of the prefactors $\alpha_i$, which depend on
$\Lambda^{(4)}_{\bar{MS} }/m_s(2 GeV)^2$, the parameter associated
with isospin breaking effects $\Omega_{IB}$ and the top mass.
Uncertainties on all these quantities make the theoretical prediction
of the SM contribution to the parameter $\varepsilon'/\varepsilon$ to
lie in a rather wide range from $5 \times 10^{-4}$ to $4
\times10^{-3}$. Therefore, the SM agrees with the observed value but
opens the possibility of sizable non-standard contribution to
$\varepsilon'/\varepsilon$.  In the following we explore the option
that non-universal A terms can account for the observed direct CP
asymmetry in kaon decays.

The dominant gluino contributions to the direct CP violation in kaon decays
can be safely approximated by the chromomagnetic operator,
\be
Re\left( \frac{\varepsilon'}{\varepsilon} \right)^{\tilde{g}} \approx 3\times 10^{6} \, Im \{C_g -\tilde{C}_g\}
\ee 
where $C_g$ is the Wilson coefficient corresponding to the
chromomagnetic operator $O_g$, and $\tilde{C}_g$ is obtained from
$C_g$ by changing $L \leftrightarrow R$. The leading part comes from the $LR$ mass
insertion, which is enhanced by a factor $m_{\tilde{g}}/m_s$. In this
limit, neglecting the $LL$ and $RR$ mass insertion effects, we have
\be
C_g \sim \frac{\alpha_s \pi}{m_{\tilde{q}}^2} \, \frac{m_{\tilde{g} } }{m_s} \, (\delta_{12}^d)_{LR} \ .
\ee

Due to the fact that the
relevant mass insertion is $Im\{(\delta_{12}^d)_{RL}\}$ is
proportional to 
$a^{d}-b^{d}$, we find that the parameter $\varepsilon'/\varepsilon$ is insensitive
to the value of the parameter $c^d$ and depends only on the
combination $a^{d}-b^{d}$. In Fig. (\ref{fig2}) we plot the SUSY contributions
(including the complete gluino and chargino loops) to the parameter
$\varepsilon'/\varepsilon$ versus the relevant combination $\delta
A\equiv a^{(d)}-b^{(d)}$, for $\tan\beta=5$, $m_0=200$ GeV and
$m_{1/2}=200,300$ GeV. In this class of models where $A_d \propto
m_0$, the value of $\varepsilon'/\varepsilon$ slightly depends on $m_0$.
\begin{figure}[ht!]
\begin{center}
\hspace*{-7mm}
\epsfig{file=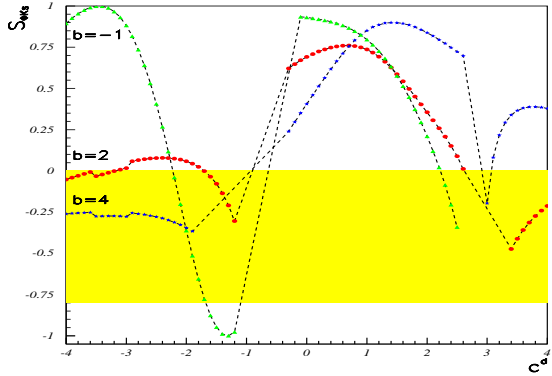,width=.5\textwidth,height=.25\textheight}\\
\caption{\footnotesize{The CP asymmetry 
$S_{\phi}$ as function of the parameter $c^d$ for $a^d=b^d-1.5$ and $b^d=4,2, -1$ and 
$\tan\beta=5$, $m_0=m_{1/2}=200$ GeV and $\delta_{12}=0$. 
The discontinuities  correspond to the regions constrained by the bound
on BR$(b \to s \gamma)$.
The yellow band corresponds to the experimental 1$\sigma$ uncertainty. }  }
\label{fig3}
\end{center}
\end{figure}
 
Now we turn to the CP violation in the $B$-meson decays. As mentioned
before, 
the recent measurements by BaBar and Belle for $S_{\phi K_S}$ evidence a deviation from the SM
prediction of  $2.7 \sigma$. Here we argue that this discrepancy can be entirely
explained in SUSY models with new sources of flavour violation and
with no need of any
further CP violating source besides $\delta_{\mathrm{CKM}}$. 
Following the parametrization of the SM and SUSY amplitudes given in .\cite{Emi},
$S_{\phi K_S}$ can be written as
\begin{eqnarray}
S_{\phi K_S}=\Frac{\sin 2 \beta +2 R_{\phi}
\cos \delta_{12} \sin(\theta_{\phi} + 2 \beta) +
R_{\phi}^2 \sin (2 \theta_{\phi} + 2 \beta)}{1+ 2 R_{\phi}
\cos \delta_{12} \cos\theta_{\phi} +R_{\phi}^2}
\end{eqnarray}
where $ R_{\phi}= \vert A^{\susy}/A^{\sm}\vert$, $\theta_{\phi}= 
\mathrm{arg}(A^{\susy}/A^{\sm})$, and in the following we set the strong phase, 
 $\delta_{12}$, to be 
zero. As it is emphasized
in Ref.\cite{Emi}, the gluino contributions can accommodate the experimental 
results of $S_{\phi K_S}$ if the magnitude of the mass insertion 
$(\delta^d_{LR(RL)})_{32}$ is 
of order $10^{-3}$ and its phase is of order one. In our model with non-universal 
$A$-terms, these values can be reached and the gluino contribution is
found to be the dominant one.
Furthermore, since we have a well defined model, our numerical
analysis is not  based on the  mass insertion approximation
but we will use the full one loop computation of the Wilson
coefficients \cite{Harnik:2002vs}. We also  
impose the constraints from the $b \to s \gamma$ branching ratio and 
$B-\bar{B}$ mixing. 

Our result for $S_{\phi K_S}$ is given in Fig. 3, where we plot the CP asymmetry
as a function of the trilinear parameter $c^d$. As discussed in the previous section,
the value of $\varepsilon'/\varepsilon$ depends only on the difference $b^d -a^d$
and it is insensitive to the value of $c^d$, which is quite relevant
for $S_{\phi K_S}$. Hence, 
we could easily find suitable values of the trilinear couplings that 
satisfy simultaneously all the meson decay observables. 
In this plot, we have fixed $b^d -a^d \simeq 1.5 $ which leads to
$\varepsilon'/\varepsilon$  of order
$1.8 \times 10^{-3}$ and considered three representative values for $b^d$ and $a^d$.
As can be seen from the figure, positive values of $b^d$ are favored
by predicting $S_{\phi K_S}$ within the experimental range. The sign of
$b^d$ depends on the relative sign of the strong phase,
$\delta_{12}$. In the case that we have considered $\delta_{12}$ is zero, hence positive values of 
$b^d$ can be regarded as natural.

\section{{\bf{Conclusions }}}

We have shown that, in supersymmetric models with non minimal flavour
structures, 
the SM phase $\delta_{\mathrm{CKM}}$ could be the only source of CP
violation and could naturally account for all the CP violating 
measurements. We have computed the baryon asymmetry generated during 
the electroweak phase transition and the parameters corresponding to 
the CP violation in meson decays, proving that our proposal for real 
and non-universal soft SUSY breaking terms provides a natural explanation for 
all of them. 

As it is well known, any new source of CP phases in the supersymmetric 
theories induces  large contributions to the EDMs which exceed their
experimental limits. We argued that SUSY models with new sources of 
flavour structures, rather than additional sources of CP violation,
can readily 
overcome this problem and, at the same time, explain the possible discrepancy between
 the SM results and
the experimental measurements of the BAU, $\varepsilon'/\varepsilon$ and $S_{\phi K_S}$. 
We have focused on the case of non-universal $A$-terms 
(which is quite natural in many SUSY models), however for non-universal 
squark masses the same conclusion can be reached.

%


\end{document}